\documentstyle[preprint,tighten,aps,floats,epsfig]{revtex}

\begin{document}
\draft
\title{
One-loop renormalization of fermionic currents with the overlap-Dirac operator}
\author{C. Alexandrou$^a$, E. Follana$^a$, H. Panagopoulos$^a$, E. Vicari$^b$}
\address{
$^a$Department of Physics, University of Cyprus,
P.O.Box 20537, Nicosia CY-1678, Cyprus.}
\address{
$^b$Dipartimento di Fisica dell'Universit\`a 
and I.N.F.N., 
Via Buonarroti 2, I-56127 Pisa, Italy.}

\date{\today}

\maketitle

\begin{abstract}

We compute the one-loop lattice renormalization of the two-quark operators
$\bar{\psi} \Gamma \psi$, where $\Gamma$ denotes the generic Dirac matrix,
for  the lattice formulation of QCD using the overlap-Dirac operator.

We also study the renormalization of quark bilinears which are more
extended and have better chiral properties.

Finally, we present improved estimates of these renormalization
constants, coming from cactus resummation and from mean field
perturbation theory.


\end{abstract}

\newpage

% ========================= BODY =========================
%\narrowtext

\section{Introduction}
\label{introduction}

Recent developments in lattice QCD have shown that chiral symmetry can be
realized on the lattice without fermion doubling (for recent reviews see e.g. Refs.\cite{N-rev-99,L-rev-99}), circumventing
the Nielsen-Ninomiya theorem~\cite{N-N-81}.
This has been achieved by introducing an overlap-Dirac operator~\cite{Neuberger-98} derived 
from the overlap formulation of chiral fermions~\cite{N-N-95}.
The simplest such example, for a massless fermion, is given by
the Neuberger-Dirac operator~\cite{Neuberger-98}
\begin{eqnarray}
D_{\rm N} &=& {1\over a} \,\rho \left[  1 + X (X^\dagger X)^{-1/2} \right],\\
X &=& D_{\rm W} - {1\over a}\rho,
\label{Nop}
\end{eqnarray}
where $D_{\rm W}$ is the Wilson-Dirac operator (with the Wilson 
parameter $r$ set to its standard value, $r=1$)
\begin{equation}
D_{\rm W} = {1\over 2} \left[ \gamma_\mu \left( \nabla_\mu^*+\nabla_\mu\right)
 - a\nabla_\mu^*\nabla_\mu \right],
\end{equation}
\begin{equation}
\nabla_\mu\psi(x) = {1\over a} \left[ U(x,\mu) \psi(x + a\hat{\mu})
- \psi(x)\right],
\end{equation}
$a$ is the lattice spacing and $\rho$ is a real parameter subject to the constraint 
$0< \rho < 2$. Nonperturbatively one expects
$-m_c< \rho < 2$, where $m_c<0$ is the critical mass associated
with the Wilson-Dirac operator.

$D_{\rm N}$ satisfies 
the Ginsparg-Wilson relation~\cite{G-W-82}
\begin{equation}
\gamma_5 D + D\gamma_5 = a D\gamma_5 D,
\label{GWrel}
\end{equation}
which protects the quark masses from 
additive renormalization~\cite{Neuberger-98,Hasenfratz-98}.
Lattice gauge theories with Ginsparg-Wilson fermions have been proved to be  renormalizable to all orders 
of perturbation theory~\cite{R-R-99}.
The Ginsparg-Wilson relation allows us to write,
at finite lattice spacing, relations that are 
essentially equivalent to those holding in the low-energy
phenomenology associated with chiral symmetry
(see e.g. Refs.~\cite{Chandrasekharan-99,K-Y-99-2}).
It indeed  implies the existence of an exact chiral symmetry of the lattice action
under the transformation~\cite{Luscher-98}
\begin{equation}
\delta_\epsilon \psi(x) = \epsilon \gamma_5 ( 1 - a D)\psi(x)  ,
\qquad\qquad \delta_\epsilon \bar{\psi}(x) = \bar{\psi}(x)\gamma_5 \epsilon,
\label{latchsym}
\end{equation}
which can be extended to the flavour non-singlet 
case\footnote{
Actually Ginsparg-Wilson fermions are invariant under a more general transformation
\begin{equation}
\delta_\epsilon \psi(x) = \epsilon \gamma_5 ( 1 - v a D)\psi(x),
\qquad\qquad \delta_\epsilon \bar{\psi}(x) = \bar{\psi}(x)\left[ 1-(1-v)a D\right]\gamma_5 \epsilon,
\label{latchsym2}
\end{equation}
where $v$ can be chosen arbitrarily. All these symmetries are essentially equivalent, leading to 
equivalent Ward identities~\cite{K-Y-99-2}.}. 
The axial anomaly then arises from the non-invariance of the
fermion integral measure under flavour-singlet chiral 
transformations \cite{Luscher-98,Luscher-99,Chiu-99,Fujikawa-99,Suzuki-99,Adams-98}.
A lattice formulation of QCD satisfying the Ginsparg-Wilson relation
overcomes the complications of the standard approach (e.g. Wilson fermions), where chiral symmetry is
violated at the scale of the lattice spacing.
The chiral symmetry ensures that the hadron
masses are free of $O(a)$ discretization errors. Indeed their
leading scaling corrections are $O(a^2)$.
The important point is that lattice Dirac operators satisfying Eq.~(\ref{GWrel}) 
are not affected by the Nielsen-Ninomiya theorem~\cite{N-N-81},
thus they need not suffer from fermion doubling.

Indeed, $D_{\rm N}$ avoids fermion doubling. 
The locality properties of $D_{\rm N}$ in the presence of a gauge field are not obvious.
$D_{\rm N}$ is not strictly local, but
locality should be recovered in a more general sense, i.e. allowing 
an exponential decay of the kernel of $D_{\rm N}$ with a rate which scales
with the lattice spacing and not with the physical quantities~\cite{H-J-L-99}.
Thus, the Neuberger-Dirac operator seems to have all the right
properties that a lattice Dirac operator should have in order to describe massless quarks. 
However, its complexity renders its numerical implementation very demanding.
In this respect some progress has been achieved (see, e.g., 
Refs.~\cite{H-J-L-99,Neuberger-98-2,E-H-N-99,E-H-N-99-2,Neuberger-98-3,Chiu-99b,Neuberger-99,%
E-H-N-99-3,H-J-L-00,B-H-E-N-99,H-J-L-00b}),
and Monte Carlo simulations seem to be already feasible, at least in quenched approximation.

The matrix elements of the fermionic currents are employed to predict hadronic decay constants, electromagnetic
and weak form factors, quark masses, etc.
A knowledge of their lattice renormalization constants is necessary to relate  the matrix elements
computed using lattice simulations to the corresponding ones defined
using continuous renormalization schemes in experimental data analysis.
In this paper we compute, to one loop in perturbation theory, 
the renormalization constants of local bilinear quark operators
$\bar{\psi} \Gamma \psi$, where $\Gamma$ denotes the generic Dirac matrix,
in the lattice formulation of QCD using the Neuberger-Dirac operator.
We also extend our computation to quark bilinears which are more
extended and have improved chiral properties.
In view of the better chiral behaviour of the Neuberger-Dirac
operator, we expect it to have a wider use 
in Monte Carlo studies of hadronic physics; such
studies require knowledge of the above renormalization constants.

Perturbative computations are much more cumbersome than in the case of
Wilson fermions, due to the more complicated structure of the 
Neuberger-Dirac operator \cite{K-Y-99,A-P-V-00,I-K-N-Y-00}. 
One-loop calculations are already quite complicated, and 
require the use of symbolic manipulation packages.
We have developed such a package in Mathematica. 
An example of one-loop calculation using the Neuberger-Dirac operator is reported in Ref.~\cite{A-P-V-00},
where we computed the ratio  $\Lambda_L/\Lambda_{\overline{MS}}$
between the $\Lambda$-parameters of the lattice formulation and 
of the $\overline{\rm MS}$ renormalization scheme.

Nonperturbative methods to estimate the renormalization constants, such as those of Refs.~\cite{BMMRT-85,MPSTV-95,LSSW-97}
would in general be preferable to approximations based on perturbative calculations, due
to their better controlled systematic errors
($O(a)$ against $O(g_0^n)$). However, 
perturbative estimates may still be quite useful. They indeed provide important consistency checks.
Further, in those cases where nonperturbative methods
are very costly to implement,
as in the case of the Neuberger formulation of the Dirac operator,
perturbative methods may remain the only source of quantitative information.

One could, of course, define exactly conserved vector and axial currents having
$Z_A=Z_V=1$, following Noether's  procedure, (i.e. performing
variations of the action with respect to the non-singlet chiral
transformations of Eq.~(\ref{latchsym})).
However, these Noether currents are rather complicated extended
objects, which might be cumbersome to use in actual simulations.
Their expressions for the Neuberger-Dirac operator can be found in Ref.~\cite{K-Y-99-2}.

The paper is organized as follows:
In Sec.~\ref{sec2} we report the one-loop calculations necessary for the evaluation of  the lattice
renormalizations of the two-quark operators. 
In Sec.~\ref{sec3} we improve the one-loop estimates by performing a
resummation to all orders of a certain
class of gauge invariant diagrams, dubbed ``cactus''
diagrams~\cite{cactus1,cactus2,cactus3}.  
We also compare them with the results of other improvement recipes~\cite{Parisi-81,L-M-93},
such as the so-called tadpole improved perturbation theory.

\section{One-loop lattice renormalization of the two-quark operators}
\label{sec2}

\subsection{Formulation of the problem}
\label{sec2a}

The lattice regularization of QCD we consider is described by the action
\begin{equation}
S_{\rm L} =  {1\over g_0^2} \sum_{x,\mu,\nu}
{\rm Tr}\left[ 1 - U_{\mu\nu}(x) \right]  +
 \sum_{i=1}^{N_f} \sum_{x,y} \bar{\psi}_i(x)D_{\rm N}(x,y)\psi_i(y).
\label{latact}
\end{equation}
where $U_{\mu\nu}(x)$ is the usual product of $SU(N)$ link variables
$U_{\mu}(x)$ along the perimeter of a plaquette
originating at $x$ in the positive $\mu$-$\nu$ directions,
and $N_f$ is the number of massless quark flavours.

The observables we study are bilinear quark operators of the form
\begin{equation}
O_i = \bar{\psi}(x) \Gamma_i \psi(x),
\label{operators}
\end{equation}
where $\Gamma_i$ denotes generic Dirac matrices, i.e.
\begin{equation}
1,\,\gamma_5,\,\gamma_\mu,\,\gamma_\mu\gamma_5,\,\sigma_{\mu\nu}\gamma_5.
\label{diracmatrices}
\end{equation}
Specific bilinear operators are denoted according to their Lorentz group
transformations: 
$S(x) \equiv \bar{\psi}(x)\psi(x)$,
$P(x) \equiv \bar{\psi}(x)\gamma_5 \psi(x)$,
$V_\mu(x) \equiv \bar{\psi}(x)\gamma_\mu \psi(x)$,
$A_\mu(x) \equiv \bar{\psi}(x)\gamma_\mu\gamma_5\psi(x)$, and
$T_{\mu\nu}(x) \equiv \bar{\psi}(x)\sigma_{\mu\nu}\gamma_5 \psi(x)$.
In the above definitions, flavour indices are left unspecified.
Indeed, this is sufficient for our purpose, 
since one-loop calculations are unaffected.

The lattice renormalization constants $Z_{O_i}$ of the operators $O_i$ are
defined once a renormalization scheme is given: 
we will consider the $\overline{\rm MS}$ renormalization scheme.
In order to simplify our calculations we work with zero quark
mass. This is justified 
by the fact that renormalization constants in the
$\overline{\rm MS}$ scheme are independent of the fermionic mass.

We also introduce the following bilinear operators 
\begin{eqnarray}
&S'\equiv \bar{\psi} \left( 1 - \case{1}{2} a D_{\rm N}\right) \psi,
\qquad &P'\equiv \bar{\psi} \gamma_5 \left( 1 - \case{1}{2} a D_{\rm N}\right) \psi,\\
&V'_\mu \equiv \bar{\psi} \gamma_\mu \left( 1 - \case{1}{2} a D_{\rm N}\right) \psi,
\qquad &A'_\mu \equiv \bar{\psi} \gamma_\mu \gamma_5 \left( 1 - \case{1}{2} a D_{\rm N}\right) \psi,
\label{op}
\end{eqnarray}
which are extended, and local only in the more general sense holding for $D_{\rm N}$.
They are interesting since 
under the lattice chiral symmetry of Eq.~(\ref{latchsym}) they transform as the corresponding operators
in the continuum. Indeed one can easily check that
\begin{equation}
\delta_\epsilon  S' = 2 P'\epsilon,\qquad\qquad \delta_\epsilon P' = 2S'\epsilon,
\label{trspp}
\end{equation}
and under the non-singlet transformations 
\begin{equation}
\delta_{\epsilon^a} \psi_i = \epsilon^a T^a_{ij} \gamma_5 ( 1 - a D)\psi_j,\qquad\qquad
\delta_{\epsilon^a} \bar{\psi}_i = \bar{\psi}_j \gamma_5 T^a_{ji}\epsilon^a,
\end{equation}
the vector and axial currents $V'$ and $A'$ transform as
\begin{equation}
\delta_{\epsilon^b} V'^a = if^{abc} \epsilon^b A'^c,\qquad\qquad 
\delta_{\epsilon^b} A'^a = if^{abc} \epsilon^b V'^c.
\end{equation}
Using the corresponding Ward identities one can show that  
\begin{equation}
Z_{S'}=Z_{P'},\qquad\qquad Z_{V'}=Z_{A'}.
\label{zpr}
\end{equation}
We mention that, as pointed out in Refs.~\cite{Chandrasekharan-99,K-Y-99-2},
the spontaneous breaking of the  non-singlet chiral symmetry, Eq.~(\ref{latchsym}), is related,
for finite lattice spacing, to the vacuum expectation value (condensate) of the operator $S'$,
i.e. it occurs if $\langle S' \rangle \neq 0$.

One can also prove that the renormalization constants of 
$O'_i \equiv \bar{\psi}\Gamma_i (1 - \case{1}{2}aD_N) \psi$ coincide
with those of the corresponding operators
$O_i\equiv \bar{\psi}\Gamma_i \psi$. 
This can be seen from the path integral representation for 
$Z_{O_i}$: Upon fermionic integration, the factor of $D_N$ in the
definition of $O'_i$ will cancel against a fermion propagator; there
will remain another propagator which will give a vanishing
contribution to $Z_{O_i}$, since it cannot develop any ${\cal O}(1/a)$
singularities (by virtue of chiral invariance, which forbids additive
mass renormalization).
Thus, using the relations Eq.~(\ref{zpr}), we also have that 
\begin{equation}
Z_{S}=Z_{P},\qquad\qquad Z_{V}=Z_{A}.
\label{ZZZ}
\end{equation}

Note that if one introduces the bare fermionic mass by writing the Dirac
operator as in Ref.~\cite{Niedermayer-99} 
\begin{equation}
\left(1 - \case{1}{2} a m_0 \right) D_{\rm N} + m_0,
\end{equation}
then the renormalization of the fermionic mass is related to
that of the operator 
$\bar{\psi} \left( 1 - \case{1}{2} a D_{\rm N}\right) \psi$, denoted
$Z_S$, by the relation 
\begin{equation}
Z_m = Z^{-1}_S.
\end{equation}

It is worth mentioning that the chiral symmetry of the Neuberger-Dirac operator
ensures the absence of $O(a)$ discretization errors only for the spectrum
of the theory. 
In order to achieve the same property for generic matrix elements of
local operators, 
one should employ improved operators~\cite{H-M-P-R-S-91}.
This issue is discussed in Ref.~\cite{C-G-H-R-S-99} for bilinear
operators constructed with Ginsparg-Wilson fermions; there it is shown that,
for massless quarks,  such
an improvement can be achieved using the operators
\begin{equation}
O''_i \equiv \bar{\psi} \left( 1 - \case{1}{2} a D_{\rm N}\right) \Gamma_i
\left( 1 - \case{1}{2} a D_{\rm N}\right) \psi.
\label{iop}
\end{equation}
Proceeding as before, one can prove 
that the renormalization constants of the operators
$O''_i$
are the same as those of the corresponding $O_i$.

\bigskip
The standard perturbative computation of the lattice renormalizations $Z_{O_i}$
of the operators $O_i$ requires the calculation of the 
two-quark one-particle irreducible function, $\Gamma(p)$, and the 
two-quark one-particle irreducible functions $\Gamma_{O_i}(p)$
with an insertion of the operators $O_i$.
In the massless case,
the quark-field renormalization $Z_\psi$ is obtained through the relation
\begin{equation}
\Gamma^{\overline{\rm MS}}(p) = Z_\psi(a\mu) \Gamma^{\rm L}(p)
\label{zpsi}
\end{equation}
where $\Gamma^{\rm L}(p)$ is the two-quark function calculated on the lattice
(in the limit $a\rightarrow 0$), and
$\Gamma^{\overline{\rm MS}}(p)$ is the $\overline{\rm MS}$-renormalized
two-point function, which can be computed in the continuum.
Indeed a simple calculation gives
\begin{equation}
\Gamma^{\overline{\rm MS}}(p) = i p_\mu \gamma_\mu \left[ 1 + g^2 c_F
\left( c \ln \frac{\mu^2}{p^2} + b^{\overline{\rm MS}} \right) + O(g^4)\right],
\label{twopms}
\end{equation}
where 
\begin{equation}
c={\alpha\over 16\pi^2} ,\qquad\qquad b^{\overline{\rm MS}}={\alpha\over 16\pi^2}, 
\label{contcoeffs}
\end{equation}
and $\alpha$ is the gauge parameter.
The renormalizations $Z_{O_i}$ can be obtained by the equation
\begin{equation}
\Gamma_{O_i}^{\overline{\rm MS}}(p) = Z_{O_i}(a\mu) Z_\psi(a\mu) \Gamma_{O_i}^{\rm L}(p)
\label{zo}
\end{equation}
where $\Gamma_{O_i}^{\rm L}(p)$ and 
$\Gamma_{O_i}^{\overline{\rm MS}}(p)$ are the two-quark functions with an insertion of $O_i$
calculated respectively on the lattice and in the continuum using the $\overline{\rm MS}$ renormalization scheme.
Setting
\begin{equation}
\Gamma_{O_i}^{\overline{\rm MS},{\rm L}}(p) = \Gamma_i B_{O_i}^{\overline{\rm MS},{\rm L}}(p),
\label{Bfunc}
\end{equation}
one has
\begin{equation}
B_{O_i}^{\overline{\rm MS}}(p) = 
1 + g^2 c_F
\left( c_{O_i}\ln \frac{\mu^2}{p^2} + b^{\overline{\rm MS}}_{O_i} \right) + O(g^4),
\label{cont}
\end{equation}
where (setting $\alpha=1$)
\begin{equation}
c_{S,P} = {1\over 4\pi^2}, \qquad c_{V,A} = {1\over 16\pi^2}, \qquad c_{T} = 0,
\label{covalues}
\end{equation}
and
\begin{equation}
b^{\overline{\rm MS}}_{S,P}= {3\over 8\pi^2},\qquad 
b^{\overline{\rm MS}}_{V,A} = {1\over 16\pi^2},\qquad
b^{\overline{\rm MS}}_{T} = 0.
\label{bmsvalues}
\end{equation}
In the above formulae we have neglected terms proportional to $p_\mu p_\nu/p^2$;
these are present also in the lattice expressions, and
cancel out in the calculation of the renormalizations.

On the lattice one finds
\begin{equation}
\Gamma^{\rm L}(p) = i p_\mu \gamma_\mu \left[ 1 + g^2 c_F
\left( - c \ln a^2p^2 + b^{\rm L} \right) + O(g^4)\right],
\label{twoplatl}
\end{equation}
and
\begin{equation}
B_{O_i}^{\rm L}(p) = 
1 + g^2 c_F
\left( -c_{O_i}\ln a^2 p^2 + b^{\rm L}_{O_i} \right) + O(g^4).
\label{latt}
\end{equation}
Using Eqs.~(\ref{zpsi}) and (\ref{zo}), we obtain
\begin{eqnarray}
Z_\psi &=& 1 + g^2 c_F
\left( c \ln a^2 \mu^2 + b^{\overline{\rm MS}}-b^{\rm L} \right) + O(g^4),\label{zpsil}\\
Z_{O_i} &=& 1 + g^2 c_F
\left[ \left( c_{O_i} - c\right) \ln a^2 \mu^2 +
b^{\overline{\rm MS}}_{O_i} - b^{\overline{\rm MS}}-b^{\rm L}_{O_i} + b^{\rm L} \right] + O(g^4),\label{zol}
\end{eqnarray}
Note that $Z_{O_i}$ are independent of the gauge parameter $\alpha$.

So, to compute $Z_{O_i}$ we need to evaluate the constants
$b^{\rm L}$ and $b^{\rm L}_{O_i}$ by a one-loop perturbative calculation on the lattice.
We have to expand the Neuberger-Dirac operator in powers
of $g_0$, and use the resulting vertices to construct
the diagrams related to the one-particle irreducible functions $\Gamma^{\rm L}(p)$
and $\Gamma^{\rm L}_{O_i}(p)$. In the appendix we list the relevant formulae for our
one-loop calculations.

\subsection{Results and discussion}
\label{sec2b}

Two diagrams contribute to $\Gamma^{\rm L}(p)$, shown
in Figure 1; for $B^{\rm L}_{O_i}(p)$, there is only one diagram to
one-loop, shown in Figure 2. Given that the 4-point vertex contains a
part with an internal 
momentum ($k$ in Eq.~(\ref{vertices})), the corresponding part of the second 
diagram in Fig. 1 actually has the same connectivity as the first diagram.
\\

\begin{figure}[h]
\begin{center}\mbox{\epsfysize=10cm\epsfxsize=10cm\epsffile{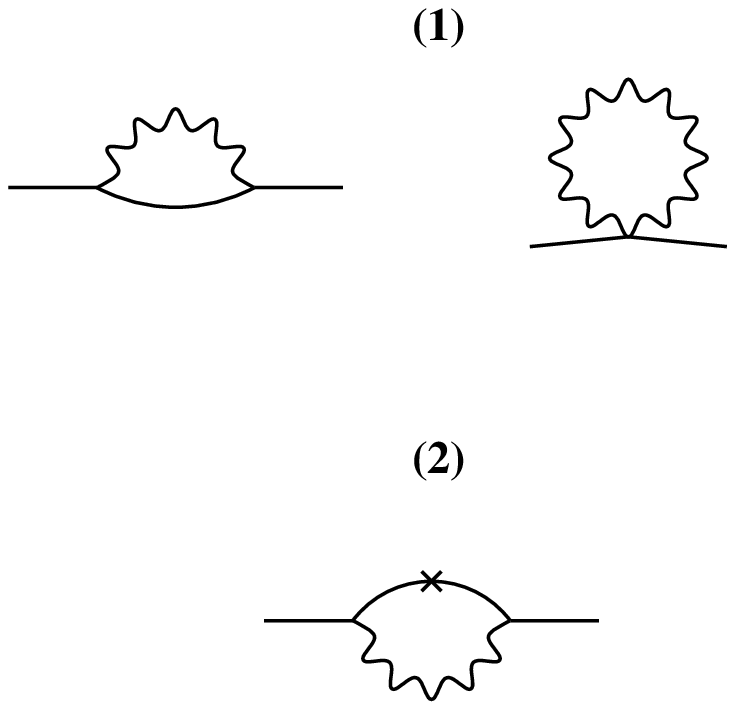}}
\qquad\qquad\qquad\qquad
\end{center}

\small FIGURES 1, 2.\ \ One-loop diagrams contributing to the functions $\Gamma^{\rm L}(p)$
(Fig. 1), and $B^{\rm L}_{O_i}(p)$ (Fig. 2).
Wavy (solid) lines represent gluons (fermions); a cross
is an insertion of the operator $O_i\,$.
\label{fig1and2}
\end{figure}

The algebra involving lattice quantities was performed using a
symbolic manipulation package which we have developed in Mathematica. 
For the purposes of the present work, this package was
augmented to include the propagator and vertices of the overlap action.

We express our results in the form of Eqs.~(\ref{twoplatl}),
(\ref{latt}). The parameters $b^{\rm L}, b^{\rm L}_{O_i}$ depend on
$\rho$, but not on $N$ or $N_f$.

To extract the $p$-dependence, we first isolate the 
divergent terms; these are responsible for the 
logarithms. There are only a few such terms, and in the pure
gluonic case their values are well known. We can use
these values also in diagrams with fermions, applying successive
subtractions of the type: 
\begin{equation}
{1\over \overline{q}^2} = 
{1\over \widehat{q}^2} + \left( {1\over \overline{q}^2} -
{1\over \widehat{q}^2}\right)
\end{equation}
where $\overline{q}^2$ is the 
inverse fermionic propagator.
All remaining terms now contain no divergences, and can be evaluated by Taylor
expansion in $ap$.

At this stage, one is left with expressions which no longer contain
$p$ and must be numerically integrated over the loop momentum.
Given the complicated form of the overlap vertices, these expressions turn out 
to be quite lengthy, containing a few hundred terms in the cases at hand. 

The integration is done in momentum space over finite lattices;
an extrapolation to infinite size is then performed, in the manner of 
Ref.~\cite{C-F-P-V-98}. We evaluated the integrals for a range of values of 
the parameter $\rho$, as presented in Table 1. 
For all values of $\rho$ that we quote, 
lattice sizes $L\le 128$ are sufficient to yield answers to at least 7
significant digits (the uncertainty coming from a systematic error in the 
extrapolation, which can be estimated quite accurately). As the endpoints of 
the perturbative domain of $\rho$ are approached ($\rho \to 0,\ \rho \to 2$), 
some of the quantities we calculate require increasingly 
larger lattices, for similar accuracy; this is, of course, a
reflection of the divergences in the propagator at these endpoints. 
Figure 3 shows the dependence of our results on $\rho$, for the whole
range $0<\rho<2$.

\begin{figure}[tb]
\mbox{\epsfysize=12cm\epsfxsize=16cm\epsffile{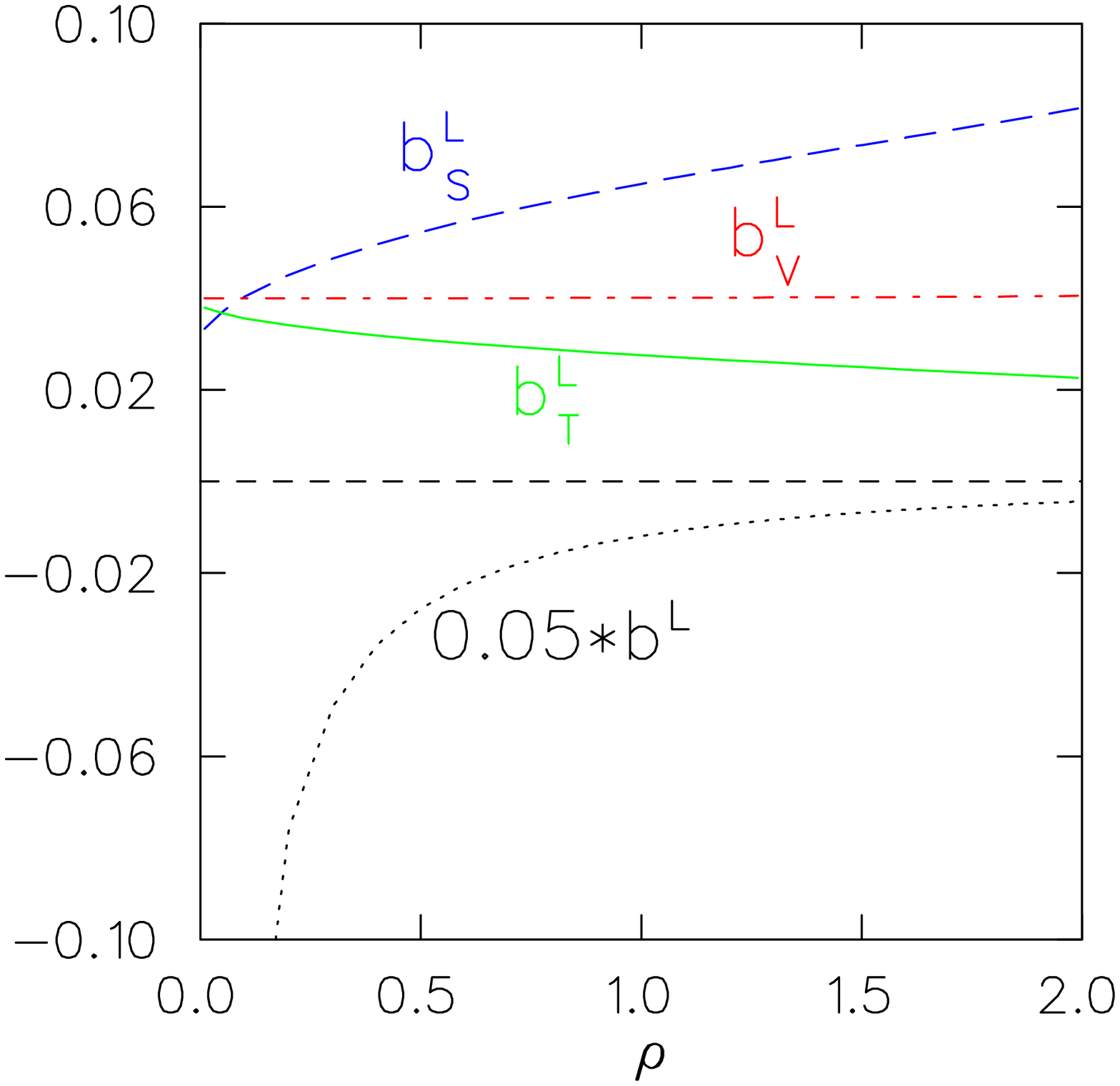}}

\begin{center}
\small FIGURE 3.\ \ The coefficients $b^{\rm L}(\rho), b^{\rm L}_{O_i}(\rho)$, as a function of the 
parameter $\rho\,$.
\end{center}
\label{fig3}
\end{figure}

\medskip
A number of consistency checks, some of which are rather nontrivial,
may be performed on our results. In particular:
\begin{itemize}
\item The logarithmic coefficients must equal those of the continuum.
\item Terms proportional to $p_\mu p_\nu / p^2$, which appear in
$Z_V,\ Z_A$, should match those of the continuum.
\item $Z_V = Z_A, \quad Z_S = Z_P\,$.
\item No ${\cal{O}}(p^0)$ terms must appear in $\Gamma^{\rm L}(p)$,
i. e. no additive mass renormalization, as required by chiral
symmetry.
\end{itemize}
Our results fulfill all of the above requirements.
The last one serves also to verify our estimates of the systematic
errors coming from the extrapolation.

\pagebreak
For quick reference, we write the one-loop values of
$Z_\psi(\alpha=1),\ Z_{O_i}$ at 
$\rho=1$, as follows:
\begin{eqnarray}
Z_\psi  &=& 1 + g^2 c_F \bigl[\, (\ln a^2 \mu^2)/16\pi^2 - 0.231966 \,\bigr] \\
Z_{S,P} &=& 1 + g^2 c_F \bigl[\, 3\, (\ln a^2 \mu^2)/16\pi^2 + 0.204977 \,\bigr] \\
Z_{A,V} &=& 1 + g^2 c_F \bigl[\, 0.198206  \,\bigr] \\
Z_{T}   &=& 1 + g^2 c_F \bigl[\, - (\ln a^2 \mu^2)/16\pi^2 + 0.204392 \,\bigr]
\end{eqnarray}

\section{Improved perturbation theory}
\label{sec3}

In order to improve the estimates coming from lattice perturbation
theory, one may perform a resummation to all orders of the so-called
``cactus'' diagrams~\cite{cactus1,cactus2,cactus3}. 
Briefly stated, these
are gauge--invariant tadpole diagrams which become disconnected if any one of their
vertices is removed. The original motivation of this procedure is the well known
observation of ``tadpole dominance'' in lattice perturbation theory.
In the following we refer to Ref.~\cite{cactus1} for 
definitions and analytical results.

Since the contribution of standard tadpole diagrams is not gauge invariant,
the class of gauge invariant diagrams we are considering needs further specification.
By the Baker-Campbell-Hausdorff (BCH) formula, the product of link variables
along the perimeter of a plaquette can be written as
\begin{eqnarray}
U_{x,\mu\nu}&& = 
e^{i g_0 A_{x,\mu}} e^{i g_0 A_{x+\mu,\nu}} e^{-i
g_0 A_{x+\nu,\mu}} e^{-i g_0 A_{x,\nu}} \nonumber \\
&&=\exp\left\{i g_0 (A_{x,\mu} + A_{x+\mu,\nu} -
A_{x+\nu,\mu} - A_{x,\nu}) + {\cal O}(g_0^2) \right\} \nonumber \\
&&=\exp\left\{ i g_0 F_{x,\mu\nu}^{(1)} +  i g_0^2
F_{x,\mu\nu}^{(2)} + {\cal O}(g_0^4)
\right\}
\end{eqnarray}
The diagrams that we propose to resum to all orders are the cactus
diagrams made of vertices containing $F_{x,\mu\nu}^{(1)}\,$.
Terms of this type come from the pure gluon  part of
the lattice action.  These diagrams dress the transverse gluon
propagator $P_A$ leading to an improved propagator $P_A^{(I)}$,
which is a multiple of the bare transverse one:
\begin{equation}
P_A^{(I)} = {P_A\over 1-w(g_0)},
\label{propdr}
\end{equation}
where the factor $w(g_0)$ will depend on $g_0$ and
$N$, but not on the momentum.
The function $w(g_0)$ can be extracted by an
appropriate algebraic equation that
 has been derived in Ref.~\cite{cactus1} and that can be easily
solved numerically; for $SU(3)$, $w(g_0)$ satisfies:
\begin{equation}
u \, e^{-u/3} \, \left[u^2 /3 - 4u +8\right]  = 2 g_0^2, \qquad 
u(g_0) \equiv {g_0^2 \over 4 (1-w(g_0))}.
\end{equation}
The vertices coming from the gluon part of the action, Eq.~(\ref{latact}),
get also dressed using a procedure similar to the one leading to Eq.~(\ref{propdr}) \cite{cactus1}.
Vertices coming from the Neuberger-Dirac
operator stay unchanged, since their definition contains no
plaquettes on which to apply the linear BCH formula.

One can apply  the resummation of cactus diagrams to the calculation of
the renormalization of lattice operators.
Approximate expressions are obtained by dressing the corresponding one-loop
calculations.
Applied to a number of cases of interest~\cite{cactus1,cactus2}, this procedure yields 
remarkable improvements when compared with the available nonperturbative estimates.
As regards numerical comparison with other improvement schemes, such as
boosted perturbation theory~\cite{Parisi-81,L-M-93}, cactus resummation fares equally well on all the
cases studied~\cite{cactus3}. It is worth mentioning in passing that
cactus resummation also affords us a systematic means of improving
perturbation theory, by successively dressing higher loop diagrams.

\medskip
Let us consider the renormalizations $Z_V, Z_A$ of isovector fermionic currents,
which are finite functions of the bare coupling $g_0$.
At one-loop order we have
\begin{equation}
Z_{V,A} = 1 + g_0^2 z_{V,A} + \ldots,
\end{equation}
where the constants $z_{V}$ and $z_{A}$ have been calculated in Sec.\ref{sec2}.
The cactus dressing of the above one-loop expressions can be simply obtained by
using the dressed transverse gluon propagator, Eq.~(\ref{propdr}).
We thus obtain the following approximate 
expressions
\begin{equation}
Z_{V,A} \approx 1 + g_0^2 {z_{V,A} \over 1-w(g_0^2)} 
\label{cadr}
\end{equation}
Let us apply the above formula to the lattice $SU(3)$ gauge theory,
for which $z_{V,A} = 0.26427$, at $g_0=1$ which is a typical value for Monte Carlo simulation.
Since~\cite{cactus1}, 
\begin{equation}
1-w(g_0=1)=0.749775,
\end{equation}
we find $Z_{V,A}(\rho=1) \simeq  1.35$

Other recipes of improvement
have been proposed in the literature 
(see e.g. \cite{L-M-93}, and \cite{C-L-V-98} for a review of them)
that essentially consist in a better choice of the expansion parameter.
Among them we mention the so-called tadpole improvement~\cite{L-M-93}
(MFI) motivated by mean-field arguments,
in which one scales the link variable with $u_0(g_0^2)\equiv 
\langle \case{1}{N} {\rm Tr}\, U_{x,\mu\nu}\rangle^{1/4}$ as measured
in the Monte Carlo simulation.
Accordingly one rescales
the coupling constant: $g_0^2\rightarrow g_{\rm mf}^2 = g_0^2/u_0^4$.
Thus, one obtains a mean-field improved expansion 
\begin{equation}
Z_{V,A} =  u_0\left[ 1 + g_{\rm mf}^2 \left( z_{V,A} + {1\over 12}\right)
+O(g_{\rm mf}^4)\right]
\label{Zvamf}
\end{equation}
For example, for $SU(3)$ in the quenched approximation and at $g_0^2=1$
one finds $u_0\simeq 0.878$ and  $g_{\rm mf}^2\simeq 1.68$. 
Putting  these number in Eq.~(\ref{Zvamf}) we obtain $Z_{V,A} \simeq 1.39$,
which is in reasonable agreement with the estimate coming using the cactus resummation.

\bigskip\bigskip
\noindent
{\bf Acknowledgements:} H. P. would like to acknowledge the warm
hospitality extended to him by the Theory Group in Pisa during various
stages of this work. H. P. and E.V. would like to thank L. Del Debbio
for useful discussions.

\appendix

\section{}
\label{appa}

In order to perform the lattice perturbative calculation 
we must formally expand $D_{\rm N}$ in powers of $g_0$. 
We list here the relevant expressions for the propagator and vertices,
following Ref.~\cite{K-Y-99}. 

Let us first write down the weak coupling expansion of
the Wilson-Dirac operator $D_{\rm W}$. This will be useful for
constructing the relevant vertices of $D_{\rm N}\,$.
We write
\begin{equation}
X(q,p) = D_{\rm W}(q,p) - {1\over a}\rho = 
X_0(p)(2\pi)^4\delta^4(q-p) + X_1(q,p) + X_2(q,p) + O(g_0^3),
\label{aqp}
\end{equation}
where
\begin{equation}
X_0(p)= {i\over a} \sum_\mu \gamma_\mu \sin a p_\mu + {1\over a} \sum_\mu
(1-\cos a p_\mu) - {1\over a} \rho,
\label{a0}
\end{equation}
\begin{eqnarray}
X_1(q,p) &=& g_0 \int d^4 k \delta(q-p-k) A_\mu(k) V_{1,\mu}(p+k/2),\label{a1}\\
V_{1,\mu}(q) &=&
i\gamma_\mu \cos aq_\mu + \sin a q_\mu,\nonumber
\end{eqnarray}
\begin{eqnarray}
X_2(q,p) &=& {g_0^2\over 2} \int {d^4 k_1 \, d^4 k_2\over (2\pi)^4}
 \delta(q-p-k_1-k_2) A_\mu(k_1)A_\mu(k_2) V_{2,\mu}(p+k_1/2+k_2/2),\label{a2}
\\
V_{2,\mu}(q) &=&
-i\gamma_\mu a\sin aq_\mu + a \cos a q_\mu.\nonumber
\end{eqnarray}

The Fourier transform of the Neuberger-Dirac operator takes the form
\begin{equation}
{1 \over \rho} D_{\rm N}(q,p) = D_0(p) (2\pi)^4\delta^4(q-p) + \Sigma(q,p).
\end{equation}
$D_0(p)$ is the tree level inverse propagator:
\begin{equation}
D_0^{-1}(p) = {-i\sum_\mu \gamma_\mu \sin ap_\mu \over 2 \left[ \omega(p) + b(p)\right] }
+ {a\over 2},
\label{d0}
\end{equation}
where
\begin{eqnarray}
\omega(p) &=& {1\over a} \left( \sum_\mu \sin^2 ap_\mu + \bigl[ 
\sum_\mu (1-\cos ap_\mu ) - \rho \bigr]^2 \right)^{1/2}, \\
b(p)&=& {1\over a} \sum_\mu (1-\cos ap_\mu) - {1\over a} \rho.
\end{eqnarray}
The function $\Sigma(q,p)$ can be expanded in powers of $g_0$ as
\begin{eqnarray}
a\Sigma(q,p) = &&
{1\over \omega(p) + \omega(q)}
\left[X_1(q,p) - {1\over \omega(p)\omega(q)} X_0(p) X^\dagger_1(p,q) X_0(q)\right] \nonumber\\
&&+
{1\over \omega(p) + \omega(q)}
\left[X_2(q,p) - {1\over \omega(p)\omega(q)} X_0(p) X^\dagger_2(p,q) X_0(q)\right] \nonumber\\
&&+
\int {d^4 k\over (2\pi)^4}
{1\over \omega(p) + \omega(q)}
{1\over \omega(p) + \omega(k)}
{1\over \omega(q) + \omega(k)}\times \nonumber \\
&&\;\;\Biggl[
-X_0(p)X_1^\dagger(p,k)X_1(k,q) 
 -X_1(p,k)X_0^\dagger(k)X_1(k,q) -X_1(p,k)X_1^\dagger(k,q)X_0(q) \nonumber \\
&&\;\;+{\omega(p)+\omega(q)+\omega(k)\over \omega(p)\omega(q)\omega(k)}
X_0(p)X_1^\dagger(p,k)X_0(k)X_1^\dagger(k,q)X_0(q)\Biggr] + ...
\label{vertices}
\end{eqnarray}
From $\Sigma(q,p)$ one can read off the vertices necessary for the
one-loop calculations presented in this paper.

\bigskip\bigskip

\begin{table}[ht]

\begin{minipage}{3cm}
\hfill
\end{minipage}
\begin{minipage}{10cm}
\caption{Coefficients $b^{\rm L}(\rho), b^{\rm L}_{O_i}(\rho)$.}
\label{tab1}
\begin{tabular}{ccccc}
{$\rho$}&$b^{\rm L}$ &$b^{\rm L}_{S,P}$ &$b^{\rm L}_{V,A}$ &$b^{\rm L}_T$\\
\tableline \hline
0.2  & 1.5236150 & 0.0450336 & 0.0399790 & 0.0340724 \\
0.3  & 0.9841497 & 0.0486865 & 0.0399888 & 0.0328679 \\
0.4  & 0.7155175 & 0.0517536 & 0.0399997 & 0.0318600 \\
0.5  & 0.5550679 & 0.0544426 & 0.0400116 & 0.0309796 \\
0.6  & 0.4486267 & 0.0568654 & 0.0400248 & 0.0301895 \\
0.7  & 0.3729964 & 0.0590910 & 0.0400393 & 0.0294670 \\
0.8  & 0.3165894 & 0.0611655 & 0.0400554 & 0.0287970 \\
0.9  & 0.2729750 & 0.0631217 & 0.0400733 & 0.0281687 \\
1.0  & 0.2382987 & 0.0649842 & 0.0400931 & 0.0275744 \\
1.1  & 0.2101107 & 0.0667722 & 0.0401153 & 0.0270079 \\
1.2  & 0.1867796 & 0.0685012 & 0.0401400 & 0.0264646 \\
1.3  & 0.1671775 & 0.0701842 & 0.0401679 & 0.0259407 \\
1.4  & 0.1504998 & 0.0718326 & 0.0401993 & 0.0254331 \\
1.5  & 0.1361573 & 0.0734568 & 0.0402349 & 0.0249392 \\
1.6  & 0.1237088 & 0.0750665 & 0.0402755 & 0.0244568 \\
1.7  & 0.1128172 & 0.0766713 & 0.0403222 & 0.0239840 \\
1.8  & 0.1032209 & 0.0782815 & 0.0403762 & 0.0235194 \\
\end{tabular}
\end{minipage}

\vspace{2cm}
\end{table}

% ========================= REFERENCES =========================

\end{document}